\documentclass[a4paper,12pt,oneside]{article}
\usepackage[latin1]{inputenc}
\usepackage{amsmath}
\usepackage{amssymb}
\usepackage{amsfonts}
\usepackage{amssymb}
\usepackage{graphicx}
\usepackage{footnote}
\usepackage{float}
\usepackage{subfigure}
\DeclareGraphicsExtensions{.bmp,.png,.pdf,.jpg,.eps}
\usepackage{physics}

\usepackage[colorlinks]{hyperref}
\hypersetup{
    colorlinks=true,
    linkcolor=blue,
    filecolor=magenta,  
    urlcolor=[rgb]{0.00,0.00,0.50},    
    citecolor=red,
    }

\usepackage{url}

\advance \textheight by \topskip
\usepackage{amsmath}
\usepackage[english]{babel}
\makeatletter
 \@addtoreset{equation}{section}
\makeatother
\usepackage[latin1]{inputenc}
\usepackage{amsmath}
\numberwithin{equation}{section}

\advance \textheight by \topskip
\usepackage{amsmath}
\usepackage[english]{babel}

\def\be{\begin{equation}}
\def\ee{\end{equation}}

\def\R{\mathbb R}

\def\C{\mathbb C}

\usepackage{todonotes}

\begin{document}

\title{{\Huge {\bf Conformal
bridge\\
[2pt]
 between asymptotic
 freedom and confinement}}}

\author{{\bf  Luis Inzunza$\,{}^a$, Mikhail S. Plyushchay$\,{}^a$, Andreas Wipf$\,{}^b$} 
 \\
[8pt]
{\small \textit{${}^a{}$Departamento de F\'{\i}sica,
Universidad de Santiago de Chile, Casilla 307, Santiago 9170124,
Chile}}\\
{\small \textit{ ${}^b{}$Theoretisch-Physikalisches Institut, Friedrich-Schiller-Universit\"at Jena, Max-Wien-Platz 1,}}
\\
{\small \textit{D-07743 Jena, Germany }}\\
[4pt]
 \sl{\small{E-mails:   
\textcolor{blue}{luis.inzunza@usach.cl},
\textcolor{blue}{mikhail.plyushchay@usach.cl},
\textcolor{blue}{wipf@tpi.uni-jena.de}
}}
}
\date{}
\maketitle

\begin{abstract}
We construct 
a nonunitary transformation
that
 relates a given ``asymptotically free" 
conformal quantum mechanical system 
$H_f$ with its  confined, harmonically trapped version  
$H_c$. 
In our construction, Jordan states corresponding to the
 zero eigenvalue  of $H_f$, 
 as well as its eigenstates and Gaussian packets
are mapped into the eigenstates, coherent states and 
squeezed states of $H_c$, respectively.
The  transformation is an 
 automorphism of the conformal $\mathfrak{sl}(2,\R)$ algebra of the nature of the 
 fourth-order root of the identity transformation,
 to which a  complex canonical transformation corresponds on the 
 classical level being  the fourth-order root of the spatial reflection.  
We investigate  the one- and two-dimensional examples
that reveal, in particular,  a curious relation
 between the two-dimensional free particle and the Landau problem.
\end{abstract}

\section{Introduction}

A well-known deficiency of the conformal quantum mechanics  is  that 
it has no invariant ground state that is annihilated by 
the
generators
of the $\mathfrak{sl}(2,\R)$ algebra, and in this sense its  conformal symmetry 
is spontaneously broken. In the context of the AdS/CFT correspondence,
this phenomenon
can be  related with a peculiar nature of a usual evolution  variable which is not a 
good global coordinate on AdS${}_2$, whose isometry is conformal 
symmetry \cite{MiStr,BlackHold1,BRFW}.
The problem emerges from the fact that 
the generators of the time translation $H$, the dilatation $D$, and the special conformal 
transformations $K$
are noncompact generators of the conformal $\mathfrak{sl}(2,\R)$ algebra, 
and the spectrum of the Hamiltonian operator $H$ of the system is 
the open interval $(0,\infty)$. 
The deficiency, however,  can be cured  from the perspective of Dirac's different forms of 
dynamics \cite{Dirac},
by considering  as a new Hamiltonian  a linear combination of these generators to be of
a compact topological nature.
 For example, one can take the operator $H+m\omega^2K$,
   which has an equidistant spectrum bounded from below, where $m$ and $\omega$
 are the mass and  frequency parameters.
 This ``regularization" was first considered by de Alfaro, Fubini and Furlan (AFF) in their 
 seminal work \cite{AFF},
 where by means of a canonical transformation of the spatial and time coordinates, 
 the conformal mechanics action describing the  asymptotically free trajectories is transformed 
 into  a modified action principle with an additional confining term in the form of a 
 harmonic trap.  The  AFF conformal mechanics model finds diverse 
 interesting applications including the black hole physics \cite{BlackHold2,BlackHold3},
 cosmology \cite{Piol,AcLi}, and holographic QCD \cite{holQCD}.
 
 A similar transformation  was considered earlier by Niederer 
 \cite{Nied}
 as the canonical transformation 
 by which the conformal-invariant classical dynamics of a free nonrelativistic particle on 
 the whole real line can be related to the dynamics of the harmonic oscillator.
A generalization of the construction allows one
 to transform the time-dependent Schr\"odinger equation of the free particle
 into that for the quantum harmonic oscillator. 
 However, a relation  between stationary states of the
 corresponding pairs of  the ``asymptotically free" and ``confined" quantum systems
 remains to be unclear.  
To establish such a  relation, we construct here a nonunitary  similarity 
 transformation 
in the form of a  ``conformal bridge between the freedom and confinement".

\vskip0.1cm
Before we pass over to the discussion of our constructions, 
it is instructive to recall  some aspects 
of the Darboux transformations  \cite{MatSal}, 
a comparison that will be useful in what follows.
\vskip0.1cm

The Darboux transformation and its generalizations 
allow one to generate new quantum systems, sometimes called superpartners, 
from a given one. 
Any generated  superpartner is
completely 
or almost completely isospectral to the initial system, and  
the energy eigenstates  of both systems are related  
to each other. 
The extended quantum system composed 
from two superpartners   is described  by 
 a  linear or nonlinear supersymmetry.
An important class of exactly solvable quantum systems 
to which Darboux transformations are applied
to produce new nontrivial systems 
includes, in particular, 
\vskip0.2cm

 (i)~the free particle, 
\vskip0.1cm

(ii) the harmonic oscillator, 
 \vskip0.1cm
 
(iii) the conformal mechanics model without confining term 
(two-particle Calogero model with omitted center of mass coordinate \cite{Calogero1}), 
\vskip0.1cm

(iv)  the AFF conformal mechanics model with a confining harmonic trap,  
\vskip0.1cm

(v) the particle in the infinite potential
well. 
\vskip0.2cm

\noindent By applying Darboux transformations to 
 the free particle (i), reflectionless quantum systems
can be obtained.   
The covariance of the Lax representation of integrable systems 
then allows one to promote  reflectionless potentials to  soliton solutions   
of  the Korteweg-de Vries (KdV) equation. 
The construction  admits a
further generalization to the case  of finite-gap solutions to the KdV
equation \cite{AraPly}.

Proceeding from the systems (ii)  and (iv), an interesting class 
of rational extensions of the harmonic oscillator and conformal
mechanics is  constructed, which have 
a discrete finite-gap type spectral structure described
by exceptional orthogonal polynomials,
and are  characterized by  an extended   deformed conformal symmetry \cite{CarPly2,CIP,LM2}.
Exceptional orthogonal polynomials also are generated
by applying  Darboux transformations 
to  the system (v) \cite{GomKam}.

Darboux transformations are usually constructed by using 
 physical or formal, nonphysical
eigenstates of the original system. Via the appropriate confluent limit of Darboux transformations,
Jordan states  of the original system enter into the construction
\cite{MatSal,confDarb,Confluent4}.
Such states are used, particularly, in 
rational deformations of the conformal mechanics with confining trap  (iv)
\cite{Klein},
and in the construction of the 
extreme type wave solutions to the complexified KdV equation 
based on the system (i) and $\mathcal{PT}$-regularized conformal mechanics (iii)
\cite{JM}.

Some pairs of the systems (i)-(v) can  be related among themselves 
by using appropriate singular Darboux transformations.  
In this way, conformal mechanics models (iii) with certain values
of the coupling constant  and model (v) 
can be generated from the system (i), while 
the AFF model (iv) can be  obtained  from 
the system (ii). 
The  systems (i) and (ii), however,
are not related by Darboux transformations.
The same is true for the pair of the systems 
(iii)  and (iv).
One may ask
whether the  quantum systems in the two indicated pairs  can be 
related by an alternative differential transformation.
\vskip0.1cm

In the present  paper we show how  the indicated systems 
can be connected 
by employing  conformal symmetry. 
\emph{A priori} it is obvious  that, as for the Darboux transformations, 
 the sought for transformations have to be 
nonunitary. We shall see, however, 
that  being similarity transformations, they
can be related to the unitary 
Bargmann-Segal transformation in the case of the pair 
of systems (i) and (ii),
where a nonunitary Weierstrass transformation
plays an essential role.
At the same time, in correspondence with the 
above indicated  modification of conformal symmetry,
the transformations
effectively  relate  Dirac's different forms
of dynamics with respect to the 
conformal $\mathfrak{sl}(2,\R)\cong \mathfrak{so}(2,1)$ symmetry. 
Also, we will observe the essential role
played by the Jordan as well as coherent and squeezed states
in our  constructions.
Our 
 ``conformal bridge transformation" can be generalized to higher dimensions, 
  and as an example we consider a  relation of the
  two-dimensional (2D) free particle system
  with the planar isotropic harmonic oscillator and the Landau problem.

\vskip0.1cm

The paper is organized as follows. 
In Sec. \ref{FreePHo}, we investigate  a relation between the quantum 
free particle and the harmonic oscillator  by 
exploiting the structure of the inverse Weierstrass transformation and the generating function 
for  Hermite polynomials  in the light of conformal symmetry. 
In Sec.  \ref{generalizationAPP} we  generalize the obtained  results 
in the context of the Dirac's different forms of dynamics.
 In  Sec. \ref{SecClas0} we explore the meaning of our 
 ``conformal bridge" transformation on 
 the classical level.
 In Sec. \ref{conformal mechanics bridge} we apply the 
  transformation to establish a relation
 between the conformal mechanics model and the AFF 
 model. This way  we generate the Schr\"odinger odd cat states 
 from eigenstates of the two-particle Calogero system. 
 In Sec.  \ref{2Dcase} the conformal bridge transformations are 
 applied to two-dimensional systems 
 to establish a relation of certain quantum states of the 
 2D free particle
 system with eigenstates and coherent states 
 of the planar isotropic harmonic  oscillator as well as the Landau 
 problem. Sec. \ref{Discussion} is devoted to the
 discussion and outlook.
Two appendixes include some  technical details.  

 \section{Free particle and harmonic oscillator}
\label{FreePHo}
In this section we investigate the transformation, based on conformal symmetry,
  by which 
the free particle and quantum harmonic oscillator systems 
can be related. 
\vskip0.1cm

{}Consider the 
generating function for Hermite polynomials $H_n(x)$,
\begin{equation}\label{Gxt}
G(x;t)=\exp(2xt-t^2)=e^{x^2}e^{-(x-t)^2}=\sum_{n=0}^\infty \frac{t^n}{n!}H_n(x)\,.
\end{equation}
{}Using it, one can obtain a chain of various  representations for $H_n(x)$\,:
\begin{eqnarray}
&H_n(x)=\frac{\partial^n}{\partial t^n}G(x;t)\vert_{t=0}=
(-1)^ne^{x^2}\frac{d^n}{dx^n}e^{-x^2}=&\nonumber\\
&\big(-e^{x^2}\frac{d}{dx}e^{-x^2}\big)^n\cdot 1=
\left(2x-\frac{d}{dx}\right)^n\cdot 1=
e^{\frac{1}{2}x^2}(x-\frac{d}{dx})^ne^{-\frac{1}{2}x^2}\,.&
\label{Hna+n}
\end{eqnarray}
Normalized  eigenfunctions of the quantum harmonic oscillator 
of mass $m$ and frequency $\omega$ described by the Hamiltonian operator
$\hat{H}_{\text{osc}}=\frac{1}{2m}\hat{p}^2+\frac{1}{2}m\omega^2 q^2$
are
\begin{equation}\label{psin(q)}
\psi_n(q)=C_n
e^{-\frac{1}{2}x^2}H_n(x)\,,\qquad
C_n=\frac{1}{\sqrt{\ell_0}}\frac{1}{\sqrt{\pi^{1/2}2^nn!}}\,,
\end{equation}
where  
 $\ell_0=\sqrt{\frac{\hbar}{m\omega}}$ 
 and $x=q/\ell_0$ is a dimensionless variable.
 Take the  units with $\hbar=m=\omega=1$,
 and introduce the  ladder operators  
 $a^+=\frac{1}{\sqrt{2}}(x-\frac{d}{dx})$, 
$a^-=(a^+)^\dagger=\frac{1}{\sqrt{2}}(x+\frac{d}{dx})$,
$[a^-,a^+]=1$. Then 
 the last equality in Eq. (\ref{Hna+n}) multiplied from the left by 
$e^{-\frac{1}{2}x^2}$ gives, up to a normalization,  the 
eigenfunctions (\ref{psin(q)}), 
that correspond  to 
the coordinate representation
of the Fock states $\vert n \rangle=\frac{(a^+)^n}{\sqrt{n!}}\vert 0\rangle$
 generated from 
 the ground state 
 of the quantum harmonic 
 oscillator.

Based on  the relation $\left(-\frac{1}{4}\frac{d^2}{dx^2} \right)
e^{2xt}= -t^2e^{2xt}$, 
one can also represent the generating function (\ref{Gxt})
as follows, 
\begin{eqnarray}
&G(x,t)=\exp\left(-\frac{1}{4}\frac{d^2}{dx^2}\right) e^{2xt}= 
\sum_{n=0}^\infty \frac{2^n t^n}{n!} 
\exp \left(-\frac{1}{4}\frac{d^2}{dx^2}\right) x^n 
\,.&\label{d2x2}
\end{eqnarray}
Comparison of (\ref{d2x2}) with (\ref{Gxt})
yields yet another representation of Hermite polynomials via
the formal inverse of the Weierstrass transform \cite{Edd,Bilo},
\begin{eqnarray}\label{Hnxn}
&H_n(x)=2^n\exp \left(-\frac{1}{4}\frac{d^2}{dx^2}\right) x^n\,.&
\end{eqnarray}
Equation  (\ref{Hnxn}) allows us   to generate the
eigenfunctions (\ref{psin(q)}) of the harmonic oscillator
by applying the  operator 
\begin{equation}\label{U0KH}
\hat{\mathfrak{S}}_0=
e^{-\hat{K}}e^{\frac{1}{2}\hat{H}_0}\,
\end{equation}
to the 
monomials  
 $x^n$, $n=0,1,\ldots$,
\begin{equation}\label{psin(x)}
\psi_n(x)=\frac{1}{\sqrt{\pi^{1/2}n!}}
\hat{\mathfrak{S}}_0
\left(\sqrt{2}x\right)^n\,.
\end{equation}

The operator (\ref{U0KH}) 
is constructed  from the operators 
\begin{eqnarray}\label{Kdef}
&\hat{H}_0=-\frac{1}{2}\frac{d^2}{dx^2}\,,\qquad
\hat{K}=\frac{1}{2}x^2\,,&
\end{eqnarray}
which together with the dilatation operator 
\begin{eqnarray}\label{Ddef}
   &\hat{D}=\frac{1}{4}(x\hat{p}+\hat{p}x)=-\frac{i}{2}\left(
   x\frac{d}{dx}+\frac{1}{2}\right)\,&
\end{eqnarray}
generate the dynamical conformal symmetry of the quantum free particle 
system,
\begin{equation}\label{slfree}
[\hat{H}_0,\hat{D}]=-i\hat{H}_0\,,
\qquad
[\hat{H}_0,\hat{K}]=-2i\hat{D}\,,
\qquad
[\hat{K},\hat{D}]=i\hat{K}\,.
\end{equation}
The operators $\hat{K}$ and $\hat{D}$  
are  the $t=0$  form of the 
explicitly depending on time
 integrals of motion
$\hat{\mathcal{K}}=\frac{1}{2}\hat{\mathcal X}^2$
and
$\hat{\mathcal{D}}=\frac{1}{4}(\hat{p}\hat{\mathcal X}+
\hat{\mathcal X}\hat{p})$, where 
$\hat{\mathcal X}=
(x-\hat{p}t)
$ is the generator of Galileo transformations of the
free particle, $\frac{d}{dt}\hat{\mathcal X}=\frac{\partial}{\partial t}\hat{\mathcal X}
-i[\hat{\mathcal X},\hat{H}_0]=0$. The set $\hat{H_0}$,  $\hat{\mathcal{D}}$
and  $\hat{\mathcal{K}}$ generates the same conformal algebra
(\ref{slfree}). Its extension by the integrals $\hat{p}$, 
$\hat{\mathcal X}$ and central element $1$ (in the chosen units)
corresponds  to the
dynamical
 Schr\"odinger symmetry of the
free particle system
\cite{NiedSch}.

In (\ref{psin(x)}), 
 $\chi_0=x^0=1$ is an eigenstate of the lowest, zero eigenvalue of 
 the free particle Hamiltonian 
 $\hat{H}_0$, while  $\chi_1=x$ is a nonphysical, 
 unbounded at infinity,
 eigenstate of $\hat{H} _0$ of the same zero eigenvalue.
 The states given by wave functions $\chi_n(x)=x^n$ with $n\geq 2$
 are the Jordan states \cite{JM}
 of the free particle corresponding to the 
 zero energy $E=0$\,:
 \begin{eqnarray}
& (\hat{H}_0)^{n}\chi_{2n}= (-\frac{1}{2})^n(2n)!\,\chi_0\,, \qquad
  (\hat{H}_0)^{n}\chi_{2n+1}= (-\frac{1}{2})^n(2n+1)!\,\chi_1\,,&\label{Jordan}\\
  &(\hat{H}_0)^{n+1}\chi_{2n}=(\hat{H}_0)^{n+1}\chi_{2n+1}=0\,,\qquad
  n=1,\ldots\,.&
  \end{eqnarray}
  The  $\chi_n(x)$ are, at the same time,  the formal eigenstates
  of the operator $2i\hat{D}$ with the same eigenvalues 
  as the  eigenfunctions   $\psi_n(x)$ of the quantum harmonic oscillator 
  Hamiltonian, 
  \begin{eqnarray}\label{Dchin}
  &2i\hat{D}\chi_n=\left(n+\frac{1}{2}\right)\chi_n\,.&
  \end{eqnarray}
  Behind the last observation lies the fact that the nonunitary operator $\hat{\mathfrak{S}}_0$ 
  intertwines generators of the 
  conformal $\mathfrak{sl}(2,\R)$ symmetry  of the quantum free particle system
  with generators of the Newton-Hooke symmetry of the quantum harmonic oscillator
  by changing the form of dynamics. 
  In details,  this can be traced out as follows.
  Using the relations $\hat{a}^-\psi_n(x)=\sqrt{n}\psi_{n-1}(x)$ and 
 $\hat{a}^+\psi_n(x)=\sqrt{n+1}\psi_{n+1}(x)$,
 we obtain from (\ref{psin(x)}) that
\begin{eqnarray}\label{sigma0xp}
 & 
 \hat{a}^-=\hat{\mathfrak{S}}_0\left(\frac{1}{\sqrt{2}}\,\frac{d}{dx}\right)
 \hat{\mathfrak{S}}_0^{-1}\,,\qquad
  \hat{a}^+=\hat{\mathfrak{S}}_0\left(\sqrt{2}x\right)\hat{\mathfrak{S}}_0^{-1}\,.&
\end{eqnarray}
 {}Therefore, the nonunitary operator
 $\hat{\mathfrak{S}}_0$
  intertwines the free particle momentum
  and coordinate operators multiplied by
  $i/\sqrt{2}$ and  $\sqrt{2}$, 
   respectively,
  with the 
  annihilation and creation operators of the quantum harmonic oscillator,
  \begin{eqnarray}\label{xdxaa}
 & 
  \hat{\mathfrak{S}}_0\left(\frac{1}{\sqrt{2}}\,\frac{d}{dx}\right)=\hat{a}^- \hat{\mathfrak{S}}_0\,,\qquad
    \hat{\mathfrak{S}}_0\left(\sqrt{2} x\right)=\hat{a}^+  \hat{\mathfrak{S}}_0\,.&
   \end{eqnarray}
Then we find that the same operator
intertwines the generators of  conformal symmetry 
of the free particle with the generators of the 
Newton-Hooke symmetry \cite{Andr} 
of the quantum harmonic oscillator,
  \begin{eqnarray}\label{Dirac2}
 & \hat{\mathfrak{S}}_0\hat{H}_0=(-2\hat{J}_-)\hat{\mathfrak{S}}_0\,,\qquad
   \hat{\mathfrak{S}}_0\hat{K}=\frac{1}{2}\hat{J}_+ \hat{\mathfrak{S}}_0\,,\qquad
   \hat{\mathfrak{S}}_0(\hat{D})=-\frac{i}{2}\hat{H}_{\text{osc}}\hat{\mathfrak{S}}_0\,,&
   \end{eqnarray}
   where
   \begin{eqnarray}\label{JJaa}
 &
\hat{J}_-=\frac{1}{2}(\hat{a}^-)^2\,,\qquad
\hat{J}_+=\frac{1}{2}(\hat{a}^+)^2\,,\qquad
\hat{H}_{\text{osc}}=\hat{a}^+\hat{a}^- +\frac{1}{2}=2\hat{J}_0\,.&
\end{eqnarray}

 In this picture, the zero energy eigenstates, $\chi_0=1$ and $\chi_1=x$
 of the free particle, together with 
 the Jordan states $\chi_n=x^n$, $n\geq 2$, 
 corresponding to  the same zero energy  (being at the same time 
 formal eigenstates
 of the noncompact $\mathfrak{sl}(2,\R)$ generator $\hat{D}=\hat{J}_2$
  multiplied by $2i$) 
 are transformed by $\hat{\mathfrak{S}}_0$  into eigenstates
 of the harmonic oscillator Hamiltonian.
 As a consequence of the
  intertwining relations (\ref{xdxaa})
 and Eqs. (\ref{Dchin}) and (\ref{Dirac2}),
 the operators $\frac{1}{\sqrt{2}}\frac{d}{dx}$ 
 and $\sqrt{2}x$ act on the described states $\chi_n(x)$, $n=0,1,\ldots$, 
  of the free particle
in the same way as the ladder operators $\hat{a}^-$ and $\hat{a}^+$ 
act  on the energy eigenstates 
of the harmonic oscillator.

The ladder operators connect
the states from the two irreducible $\mathfrak{sl}(2,\R)$ representations 
of the discrete type series
$D^+_{\alpha}$ with $\alpha=1/4$ and $\alpha=3/4$,
in which the Casimir operator 
(\ref{Casimir}) takes the same value $\hat{C}=-\alpha(\alpha-1)=3/16$,
while the compact generator $\hat{J}_0=\frac{1}{2}\hat{H}_{\text{osc}}$ has  the values
$j_{0,n}=\alpha+n$, $n=0,1,\ldots$.
These two irreducible representations of 
$\mathfrak{sl}(2,\R)$ 
are realized on subspaces spanned by 
even, $\{\psi_{2n}(x)\}$, and odd, 
$\{\psi_{2n+1}(x)\}$, $n=0,1,\ldots$, eigenstates of the
quantum harmonic oscillator. Together they
constitute an irreducible representation of the 
$\mathfrak{osp}(1\vert 2)$ superconformal algebra
with $\hat{a}^-$ and $\hat{a}^+$
as odd generators \cite{InzPlyHid}. 
The indicated separation of eigenstates of the harmonic oscillator
corresponds to a separation 
of the set of  Jordan states in (\ref{Jordan}) 
with odd and even values of the index.

The generating function (\ref{Gxt}) is  related 
with a unitary map from the 
Hilbert space $L^2(\R)$ 
to the Fock-Bargmann space \cite{Takh,Gazeau},
\begin{equation}\label{KqzGxz}
G(x,z/\sqrt{2})=\left(\pi\right)^{1/4}e^{\frac{1}{2}x^2}
U(x,z)\,.
\end{equation}
Here
\begin{equation}\label{K(q,z)}
U(x,z)=\sum_{n=0}^\infty\overline{\psi_n(x)}f_n(z)\,,
\end{equation}
with $\psi_n(x)$ being the
energy eigenstates of the harmonic oscillator in the 
coordinate representation, while
$f_n(z)=\frac{z^n}{\sqrt{n!}}$, $z\in \C$,
describe the same orthonormal states in the Fock-Bargmann 
 representation 
with scalar product 
\begin{equation}\label{FBscalar}
(f,g)=\frac{1}{\pi}\int_{\R^2}e^{-\vert z\vert^2}\, \overline{f^(z)}g(z)d^2z\,,
\qquad
d^2z=d(\text{Re}\,z)d(\text{Im}\,z)\,.
\end{equation}
The explicit form of (\ref{K(q,z)}) 
is 
\begin{eqnarray}
&U(x,z)=\pi^{-1/4}\exp\left(-\frac{1}{2}(x^2-2\sqrt{2}zx+z^2)\right)=\langle x\vert z\rangle_S\,,&
\end{eqnarray}
that corresponds to the Schr\"odinger non-normalized coherent state in the coordinate 
representation \cite{Schro}.
Equations (\ref{d2x2}) and (\ref{KqzGxz}) 
yield then the relation
\begin{equation}
U(x,z)=\hat{\mathfrak{S}}_0\phi_z(x)\,,\quad 
\text{where}\quad
\phi_z(x)=\pi^{-1/4}e^{\sqrt{2}zx}\,.
\end{equation}
The wave function $\phi_z(x)$
with $z\in\C$
corresponds to a  formal eigenstate of 
the operator $\hat{H}_0$  with eigenvalue $-z^2$,
which at pure imaginary values $z=ik/\sqrt{2}$ of $z$
is a 
plane wave eigenstate $e^{ikx}$ of the  free 
particle\footnote{Note that linear combination of the states
$\phi_z(x)$ with real $z$ are used as seed states
for Darboux  transformations
to generate soliton 
solutions to the KdV equation.
These states with complex  
$z$  are employed in the construction
of the KdV soliton Baker-Akhiezer function \cite{Diej}.}.
Therefore, in addition to relation (\ref{psin(x)})
we find that the nonunitary operator 
$\hat{\mathfrak{S}}_0$ maps (formal in the general case
of $z\in \C$) 
plane wave type eigenstates of the free 
particle
 into the Schr\"odinger 
coherent states of the quantum harmonic oscillator.
The standard, or canonical  
Schr\"odinger-Klauder-Glauber coherent states of 
the quantum harmonic oscillator in coordinate representation 
\cite{Gazeau,KlauSka,Perelomov,DeFrHu},
\begin{equation}
\langle x\vert z\rangle=e^{-\frac{1}{2}\vert z\vert^2}U(x,z):=\psi_z(x)\,,
\end{equation}
are generated from the rescaled formal plane wave eigenstates of the free particle\,:
\begin{equation}
\psi_z(x)=\hat{\mathfrak{S}}_0\left(e^{-\frac{1}{2}\vert z\vert^2}
\phi_z(x)\right)\,.
\end{equation}

Thus, the eigenstates of the quantum harmonic oscillator are
 obtained from the
Jordan states and eigenstates of the free particle
corresponding to zero energy by applying to them the nonunitary operator  $\hat{\mathfrak{S}}_0$.
These free particle's  states are also  
formal eigenstates (with purely imaginary eigenvalues) of the dilatation operator $\hat{D}$.
On the other hand, in correspondence with the first relation in
(\ref{xdxaa}), the plane wave eigenstates of the free particle, being eigenfunctions of the momentum operator, 
are mapped by $\hat{\mathfrak{S}}_0$
into coherent states of the quantum harmonic oscillator (QHO) being eigenstates of its annihilation operator.
The free particle plane wave eigenstates 
are produced then by action of the  inverse operator $\hat{\mathfrak{S}}_0^{-1}=e^{-\frac{1}{2}H_0}e^{K}$
on the coherent states of the QHO.
\vskip0.1cm

Another important class of the states of the quantum harmonic 
oscillator corresponds to the squeezed states \cite{Perelomov,Gazeau}.
The so-called single-mode squeezed states 
$\vert r\rangle=\hat{S}(r)\vert 0\rangle$
are obtained from the
ground state $\vert 0\rangle$
 by acting with the unitary operator 
 $\hat{S}(r)=\exp(r\hat{J}_--r\hat{J}_+)=\exp(-2ir\hat{J}_2)$ on it;
 see  Eqs. (\ref{JJaa}) and (\ref{Jpmdef}).
 In the coordinate representation it is  an
  infinite linear combination of eigenfunctions of the QHO,
 \be\label{squeezdedQHO}
 \psi_r(x)=\frac{1}{\sqrt{\cosh r}}\sum_{n=0}^\infty(-\tanh r)^n\frac{\sqrt{(2n)!}}{2^n n!}\psi_{2n}(x)\,.
 \ee
Using relation (\ref{psin(x)}), we find that the preimage of (\ref{squeezdedQHO})
under the action of the nonunitary operator $\hat{\mathfrak{S}}_0$ 
is given by a Gaussian  wave packet. Namely, 
\begin{eqnarray}
&\hat{\mathfrak{S}}_0\left(\frac{1}{\sqrt{\pi^{1/2}\cosh r}}\exp({-\tanh r \cdot x^2})\right)=\psi_r(x)\,.&
\end{eqnarray}
Therefore, the Gaussian wave packets of the quantum free particle system correspond 
to the single-mode squeezed states of the QHO under the similarity 
transformation generated by the nonunitary operator $\hat{\mathfrak{S}}_0$.

\section{Generalization for further applications }
\label{generalizationAPP}

The similarity transformation that allowed us to establish
a ``bridge" between the quantum free particle system and the quantum harmonic 
oscillator is given by the nonunitary operator (\ref{U0KH})
constructed from generators of the conformal algebra.
The generators of this algebra correspond to the dynamical symmetry
$\mathfrak{sl}(2,\R) \cong \mathfrak{so}(2,1) $ of 
the quantum free particle system. 
This transformation generates a change of dynamics
in the sense of Dirac \cite{Dirac}.  The initial system 
is described by  a Hamiltonian operator being 
a noncompact linear combination of the generators 
$\hat{J}_\mu$  
of the Lorentz algebra $\mathfrak{so}(2,1)$, 
$\hat{H}_0=\hat{J}_0+\hat{J}_1$.
The dynamics of the final system being  the QHO is described
by the compact generator $2\hat{J}_0=\hat{H}_{\text{osc}}=\hat{H}_0+\hat{K}$ of the same algebra.
The similarity transformation produced by $\hat{\mathfrak{S}}_0$
does not change the conformal algebra, but being
a nonunitary operator, 
it
intertwines  $\mathfrak{sl}(2,\R)$ generators 
of different topological  natures according to
Eq.  (\ref{Dirac2}). 

One can modify the similarity transformation by changing
operator (\ref{U0KH}) to
\begin{equation}\label{U0KH+}
\hat{\mathfrak{S}}=
e^{-\hat{K}}e^{\frac{1}{2}\hat{H}_0}e^{i\ln 2\cdot \hat{D}}=
e^{-\hat{K}}e^{i\ln 2\cdot \hat{D}}e^{\hat{H}_0}
\,.
\end{equation}
As a result of the inclusion of the additional unitary factor $e^{i\ln 2\cdot \hat{D}}$,
operator  (\ref{U0KH+}) acts on the coordinate and momentum $\hat{p}=-i\frac{d}{dx}$ 
operators in a more 
symmetric way. Instead of (\ref{sigma0xp}), we have 
\begin{eqnarray}\label{sigmaxp}
 & 
 \hat{\mathfrak{S}} \hat{p}
 \hat{\mathfrak{S}}^{-1}=-i\hat{a}^-\,,\qquad
\hat{\mathfrak{S}} x \hat{\mathfrak{S}}^{-1}=  \hat{a}^+\,.&
\end{eqnarray}
This is a nonunitary canonical transformation 
that is identified as the
fourth order root of the space reflection operator,  
\begin{eqnarray}
& \hat{\mathfrak{S}}: (x,\hat{p},\hat{a}^+,\hat{a}^-){\rightarrow}(\hat{a}^+,-i\hat{a}^-,-i\hat{p}, x)\,,\qquad
 \hat{\mathfrak{S}}^2: (x,\hat{p},\hat{a}^+,\hat{a}^-){\rightarrow}(-i\hat{p}, -ix, -\hat{a}^-,\hat{a}^+)\,,&\nonumber\\
& \hat{\mathfrak{S}}^4: (x,\hat{p},\hat{a}^+,\hat{a}^-){\rightarrow}(-x, -\hat{p}, -\hat{a}^+,-\hat{a}^-)\,.&
\end{eqnarray}
Therefore, from the point of view of the quantum phase space,
transformation (\ref{sigmaxp}) is the eighth order root of the identity transformation.

The operator (\ref{U0KH+}) can also be factorized as 
\begin{eqnarray}\label{SIGMA}
\hat{\mathfrak{S}}=\exp(\hat{J}_1-\hat{J}_0)\cdot \exp(i\ln 2\cdot 
\hat{J}_2)
\cdot \exp(\hat{J}_0+\hat{J}_1)
\end{eqnarray}
involving  the $\mathfrak{sl}(2,\R)$ generators $\hat{J}_\mu$.
It produces the following nonunitary automorphism 
of  $\mathfrak{sl}(2,\R)$:
 \begin{eqnarray}\label{Automor}
 & \hat{\mathfrak{S}}\hat{J}_0\hat{\mathfrak{S}}^{-1}=i\hat{J}_2\,,\qquad
   \hat{\mathfrak{S}}\hat{J}_1\hat{\mathfrak{S}}^{-1}=-\hat{J}_1\,,\qquad
   \hat{\mathfrak{S}}\hat{J}_2\hat{\mathfrak{S}}^{-1}=-i\hat{J}_0\,.&
   \end{eqnarray}
 The action of the squared nonunitary similarity transformation 
   on the  $\mathfrak{sl}(2,\R)$ generators is a rotation by $\pi$ about $J_1$,
$\hat{\mathfrak{S}}^2\,:(\hat{J}_0,\hat{J}_1,\hat{J}_2)
\rightarrow (-\hat{J}_0, \hat{J}_1,-\hat{J}_2)$,
such that  the action of $\hat{\mathfrak{S}}$ on $\hat{J}_\mu$ is identified as the
fourth order  root of the identity.

Equations (\ref{SIGMA}) and (\ref{Automor})
can be used to relate other pairs of the quantum systems 
described by different realizations of the conformal  $\mathfrak{sl}(2,\R)$ dynamics 
 in the sense of 
Dirac's different forms of 
dynamics \cite{Dirac}.

\section{Classical picture}\label{SecClas0}

In this section, we consider the classical picture
underlying the quantum transformation
based on conformal symmetry and relating 
the  quantum mechanical systems of the free particle and harmonic oscillator.
 This will allow us, in particular, to obtain an interesting reinterpretation of some aspect 
 of the quantum conformal bridge transformation
 in relation with the  unitary 
Bargmann-Segal transformation and 
the Neumann-Stone theorem.
 \vskip0.1cm

The classical analogs of the free particle's generators of conformal 
symmetry,  $H_0=\frac{1}{2}p^2$, $D=\frac{1}{2}xp$ and $K=\frac{1}{2}{x}^2$,
satisfy  the Poisson bracket relations
\begin{equation}\label{H0JD}
\{D,H_0\}=H_0\,,\qquad
 \{D,K\}=-K\,,\qquad 
 \{K,H_0\}=2D\,,
 \end{equation}
 which correspond to the quantum $\mathfrak{sl}(2,\R)\cong \mathfrak{so}(2,1)$ algebra 
 (\ref{Jmunul})
 under the identification 
 \begin{equation}\label{J0JiHKD}
 J_0=\frac{1}{2}(H_0+K)\,,\qquad
  J_1=\frac{1}{2}(H_0-K)\,,\qquad
  J_2=D\,.
  \end{equation}
The classical analog of the Casimir element (\ref{Casimir})  is presented here 
\textcolor[rgb]{0.00,0.80,0.30}{
in the form 
}  
$C=-H_0K+D^2$,
and  takes zero value, $C=0$.
Since $J_0\geq 0$, the dynamics of both  the free particle and 
the harmonic oscillator 
takes place on the upper cone surface in coordinates of
the $\mathfrak{sl}(2,\R)$ conformal algebra generators $(J_0,J_1,J_2)$.
\vskip0.1cm

The phase space functions  $H_0$, $K$ and $D$ 
generate the following canonical transformations 
of $x$ and $p$, see Appendix \ref{AppB}\,:
\begin{equation}
T_{H_0}(\alpha)(x)=x-\alpha p\,,\qquad
T_{H_0}(\alpha)(p)=p\,,
\end{equation}
\begin{equation}
T_K(\beta)(x)=x\,,\qquad
T_K(\beta)(p)=p+\beta x\,,
\end{equation}
\begin{equation}
T_D(\gamma)(x)=e^{-\frac{1}{2}\gamma}x\,,\qquad
T_D(\gamma)(p)=e^{\frac{1}{2}\gamma}p\,.
\end{equation}
Then the  composition 
\begin{equation}\label{Tabg0}
T_{\beta\alpha\gamma}:=T_K(\beta)\circ T_{H_0}(\alpha) \circ T_D(\gamma)=
 T_K(\beta)\circ T_D(\gamma) \circ T_{H_0}(2\alpha)
 \end{equation}
transforms the 
canonical variables as follows:
\begin{equation}\label{Tabg}
T_{\beta\alpha\gamma}(x)=e^{-\frac{1}{2}\gamma}\left(x(1-\alpha\beta)-\alpha p\right):=\widetilde{x}\,,\qquad
T_{\beta\alpha\gamma}(p)=e^{\frac{1}{2}\gamma}\left(p +\beta x\right):=\widetilde{p}\,.
\end{equation}
The choice 
\begin{equation}\label{abgfix}
\alpha=\frac{i}{2}\,,\qquad
 \beta=-i\,,\qquad \gamma=-\ln 2\,
\end{equation}
gives
\begin{equation}\label{tildexpa}
\widetilde{x}=a^+\,,\qquad 
\widetilde{p}=-ia^-\,,
\qquad 
\widetilde{a}^+=-ip\,,\qquad 
\widetilde{a}^-=x\,,
\end{equation}
where 
$a^-=\frac{1}{\sqrt{2}}(x+ip)$ and $a^+=\frac{1}{\sqrt{2}}(x-ip)$,
$\{a^-,a^+\}=-i$, 
are the classical analogs of the creation and annihilation  operators. 
This shows that  the transformation $T_{\beta\alpha\gamma}$ 
with the parameters fixed as in (\ref{abgfix})
is the classical analog of  the operator (\ref{U0KH+}). 

The same complex canonical transformation applied to
the generators of conformal symmetry gives 
$\widetilde{H}_0=-\frac{1}{2}(a^-)^2\,,$ 
$\widetilde{K}=\frac{1}{2}(a^+)^2\,,$ 
$\widetilde{D}=-\frac{i}{2}a^+a^-\,.$
 The $2iD$ transforms into the Hamiltonian of the classical
 harmonic oscillator, and we find that it  generates the complex  flow in
 the phase space given by  $x':=\{x,2iD\}=ix$,
 $p':=\{p,2iD\}=-ip$. Taking into account 
 relations (\ref{tildexpa}) and $2i\widetilde{D}=H_{\text{osc}}$,
 one concludes  that these relations correspond exactly 
 to the classical equations of motion for the 
harmonic oscillator, 
$\dot{a}^\pm=\{a^\pm, H_{\text{osc}}\}=\pm i a^\pm$.
\vskip0.1cm

The similarity  transformation
 (\ref{sigmaxp}) preserves the commutation relations. 
However,  
the  operator (\ref{U0KH+}) 
is not unitary with respect to the  scalar product 
\begin{equation}\label{scalq}
(\psi_1, \psi_2)=\int_{-\infty}^{+\infty}\overline{\psi_1(x)}
\psi_2(x)dx\,.
\end{equation}
This is the case since the parameters $\alpha$ and $\beta$
in the corresponding classical canonical transformation are purely imaginary,
as a result of which the transformed canonical variables
$\widetilde{x}=a^+$ and  $\widetilde{p}=-ia^-$ are complex variables,
while their quantum analogs (\ref{sigmaxp}) are not Hermitian operators
with respect to the scalar product (\ref{scalq}).
This deficiency  can be 
``cured" and reinterpreted in correspondence with the 
Stone-von Neumann theorem 
\cite{Takh}.
The transformed classical coordinate 
takes complex values, $\tilde{x}=a^+\in \C$, and its complex conjugation is 
\begin{equation}\label{x*p}
\overline{\tilde{x}}=a^-=i\tilde{p}\,.
\end{equation}
In correspondence with these properties, at the quantum level 
we pass over 
from the coordinate representation to representation in which 
$\hat{\widetilde{x}}=\hat{a}^+$ acts as the operator of multiplication 
by a complex variable $z$, $\hat{a}^+\psi(z)=z\psi(z)$,
while the operator $i\hat{\tilde{p}}=\hat{a}^-$ acts as 
 a differential
operator, $\hat{a}^-\psi(z)=\frac{d}{dz}\psi(z)$. Hence
the  quantum relation $[\hat{a}^-,\hat{a}^+]=1$ is the analog 
of the classical relation $\{\tilde{x},\tilde{p}\}=1$.
Replacing finally
the scalar product
(\ref{scalq}) by the scalar product (\ref{FBscalar}),
\begin{equation}\label{scalhol}
(\psi_1, \psi_2)=\frac{1}{\pi}\int_{\R^2}\overline{\psi_1(z)}
\psi_2(z)e^{-\bar{z}z}d^2z,\qquad 
d^2z=d(\text{Re}\,z)d(\text{Im}\,z)\,,
\end{equation}
we arrive at the Fock-Bargmann representation,
in which  the operators $\hat{a}^+=z$ and $\hat{a}^-=\frac{d}{dz}$
satisfy the relation  $(\hat{a}^+)^\dagger=\hat{a}^-$, that corresponds to the
classical identity (\ref{x*p}).
In this representation 
\begin{equation}\label{tildeH0KD+}
2i\widehat{\widetilde{D}}=\hat{H}_{\text{osc}}=\left(z\frac{d}{dz}+\frac{1}{2}\right),\qquad
\widehat{\widetilde{H}}_0=-\frac{1}{2}\frac{d^2}{dz^2}\,, \qquad
\widehat{\widetilde{K}}=\frac{1}{2}z^2\,,
\end{equation}
where $\widehat{\widetilde{D}}$ is constructed
from the phase space function $\widetilde{D}$ via an
antisymmetrization. 
In other words, the transformed operators in the Fock-Bargmann representation
can be obtained from the corresponding initial generators of conformal symmetry 
of the quantum free particle by a  formal change of $x$ to $z$.
The change of the scalar product from
(\ref{scalq}) to (\ref{scalhol}) transmutes then  the nonunitary 
similarity transformation  (\ref{sigmaxp}) into the unitary 
transformation from the coordinate to the holomorphic representation 
for the Heisenberg algebra in correspondence with 
the Neumann-Stone theorem, to which the  kernel (\ref{K(q,z)}) corresponds.


\section{Conformal mechanics bridge
}
\label{conformal mechanics bridge}

In this section, we construct in a similar way a conformal bridge  
between  the quantum  conformal mechanics systems  (iii) and (iv).
The peculiarity in comparison with the previous case of the pair of the systems
(i) and (ii) is that here the operators $\hat{x}$ and $\hat{p}$ are not observables anymore
since $\hat{p}$ is not self-adjoint  and 
the commutator of $\hat{x}$ with 
the corresponding
Hamiltonian $\hat{H}_\nu$ gives $\hat{p}$.
Thus we  work  in terms of an operator analogous to
(\ref{SIGMA}) and its associated relations (\ref{Automor}),
which are well-defined being quadratic in $\hat{x}$ and $\hat{p}$. 
\vskip0.1cm

The Hamiltonian operator of the two-body Calogero model 
with an omitted  center of mass degree of freedom is 
\begin{eqnarray}\label{Hnudef}
\hat{H}_\nu=-\frac{1}{2}\frac{d^2}{dx^2}+\Delta_\nu\,,
\qquad \Delta_\nu=\frac{\nu(\nu+1)}{2x^2}\,,
\end{eqnarray}
where
$\nu\geq -1/2$, $x\in \R^+$,  and we impose the Dirichlet
 boundary condition $\psi(0)=0$ for the wave functions.
The case $\nu=0$ corresponds to the free particle
on the half-line $\R^+$ with eigenstates 
$\psi_\kappa(x)=\sin \kappa x$ of energies $E=\frac{1}{2}\kappa^2>0$.
The limit $\lim_{\kappa\rightarrow 0}\sin \kappa x/\kappa=x$
gives a nonphysical, unbounded zero energy
eigenstate, which simultaneously is the  eigenstate of
the dilatation operator (\ref{Ddef}), $2i\hat{D}x=\frac{3}{2}x$. 
In the general case of $\nu\geq -1/2$, 
solutions of the stationary equation 
$\hat{H}_\nu \psi_{\kappa,\nu}=E_\nu(\kappa)\psi_{\kappa,\nu}$
with $E_\nu(\kappa)=\frac{1}{2}\kappa^2$, $\kappa>0$ are
\begin{equation}
\label{Calogerostates}
\psi_{\kappa,\nu}(x)=\sqrt{x}\mathcal{J}_{\nu+1/2}(\kappa x)\,,
\end{equation}
where 
\begin{equation}
\mathcal{J}_{\alpha}(\eta)=
\sum_{n=0}^{\infty}\frac{(-1)^n}{n!\Gamma(n+\alpha+1)}
\left(\frac{\eta}{2}\right)^{2n+\alpha}\,
\end{equation}
is the Bessel function of the first kind.
The solutions (\ref{Calogerostates}) include  $\nu=0$ as a
particular case.
The Hamiltonian operator $\hat{H}_\nu$ is invariant under the 
change 
 $\nu\rightarrow -(\nu+1)$, but the same transformation 
 applied to  (\ref{Calogerostates}) produces eigenfunctions
 of $\hat{H}_\nu$ not satisfying  the boundary condition 
 $\psi(0)=0$ \cite{Klein}. 
At $\nu=-1/2$, we have
$\nu=-(\nu+1)$, and the eigenstates (\ref{Calogerostates})
correspond to a particular case of a family of
self-adjoint extensions of $\hat{H}_{-1/2}$ 
\cite{Klein,FPW}.
\vskip0.1cm

Consider the set of wave functions
  $x^{\nu+1+2n}$,  $n=0,1,\ldots$.
The function with $n=0$ represents a formal, diverging at infinity,
 eigenstate of the differential operator
$\hat{H}_\nu$ with $\nu\geq -1/2$ of  eigenvalue $E=0$,
which follows from solutions  (\ref{Calogerostates})  by applying
to them the same limit as in the case of $\nu=0$.
The wave functions with $n\geq 1$   are the Jordan states of rank $n$ 
corresponding to the same eigenvalue  $E=0$ of $\hat{H}_\nu$.
The functions $x^{\nu+1+2n}$ are  at the same time 
eigenstates of the operator $2i\hat{D}$ with eigenvalues 
$\nu+2n+3/2$. 
The Jordan states with $n\geq 1$ satisfy the relations
\begin{eqnarray}
\label{j-th_element1+}
(\hat{H}_{\nu})^{j}x^{\nu+1+2n}=
\frac{(-2)^{j}\Gamma(n+1)}{\Gamma(n+1-j)}\frac{\Gamma(n+\nu+3/2)}{\Gamma(n+\nu+3/2-j)}x^{\nu+1+2(n-j)}\,,
\quad
j=0,1,\ldots,n\,,
\end{eqnarray}
which can be proved by  induction.
Equation (\ref{j-th_element1+}) extends to the case $j=n+1$ giving
$(\hat{H}_{\nu})^{n+1}x^{\nu+1+2n}=0$ due to the 
appearence of the  simple pole in the 
denominator.
\vskip0.1cm

The operator $\hat{H}_\nu$ together with the operators 
 $\hat{K}$ and $\hat{D}$ defined by Eqs. (\ref{Kdef}) and (\ref{Ddef})
 satisfy the conformal $\mathfrak{sl}(2,\R)$ algebra  
 (\ref{slfree}) as in the case of the free particle on the whole real line.
 Now we can define  the direct analog of the operator 
  $\hat{\mathfrak{S}}$  in the form\footnote{One can work, instead,
  with the analog of the operator $\hat{\mathfrak{S}}_0$, 
  but such a  change is not essential.}
  \begin{equation}\label{UKHS}
\hat{S}_\nu:=
e^{-\hat{K}}e^{\frac{1}{2}\hat{H}_\nu}e^{i\ln 2\cdot \hat{D}}=
e^{-\hat{K}}e^{i\ln 2\cdot \hat{D}}e^{\hat{H}_\nu}\,.
\end{equation}
A similarity transformation with this nonunitary operator
 produces, analogously to 
(\ref{Dirac2}), the relations
 \begin{eqnarray}\label{Dirac2+}
 & \hat{{S}}_\nu\hat{H}_\nu \hat{{S}}_\nu^{-1}=-\hat{J}_-\,,\qquad
   \hat{{S}}_\nu\hat{K}\hat{{S}}_\nu^{-1}=
   \hat{J}_+\,,\qquad
   \hat{{S}}_\nu(\hat{D})\hat{{S}}_\nu^{-1}=
   -\frac{i}{2}\hat{H}^{\text{AFF}}_\nu\,,&
   \end{eqnarray}
   where now instead of (\ref{JJaa}),  
   \begin{eqnarray}\label{JJaa+}
 &
\hat{J}_-=\frac{1}{2}(\hat{a}^-)^2-\Delta_\nu\,,\qquad
\hat{J}_+=\frac{1}{2}(\hat{a}^+)^2-\Delta_\nu\,,\qquad
2\hat{J}_0=\hat{H}^{\text{AFF}}_\nu=\hat{a}^+\hat{a}^- +\frac{1}{2}+\Delta_\nu\,.&
\end{eqnarray}
The dilatation operator multiplied by $2i$ transforms in this case
into the Hamiltonian  $\hat{H}^{\text{AFF}}_\nu$ 
of the conformal mechanics 
model of de Alfaro, \emph{et al}. \cite{AFF}.
At the same time, we note that the operators 
$x$ and $\frac{d}{dx}$ are transformed by 
the operator $\hat{{S}}_\nu$ into nonlocal operators
corresponding to the square root of the operators
$\hat{J}_+$ and $\hat{J}_-$ in (\ref{JJaa+}),
whose action violates the boundary condition $\psi(0)=0$.
\vskip0.1cm 

Application of $\hat{{S}}_\nu$ to the  states  $x^{\nu+1+2n}$
related to the system  $\hat{H}_\nu$ produces the energy eigenstates
of the AFF model, 
\begin{eqnarray}
\label{Physeigen}
&\hat{S}_\nu(\frac{1}{\sqrt{2}}x)^{\nu+1+2n}=2^{\frac{1}{4}}(-1)^n n!\psi_{\nu,n}^{\text{AFF}}(x)\,,&
\end{eqnarray}
where
\begin{eqnarray}
\label{statesLnu1}
\psi_{\nu,n}^{\text{AFF}}(x)=x^{\nu+1}\mathcal{L}_{n}^{(\nu+1/2)}(x^2)e^{-x^2/2}\,,\qquad E_{\nu,n}=2n+\nu+3/2\,
\end{eqnarray}
are the non-normalized eigenstates of the  AFF model
 and their  corresponding energy values,
and
\begin{equation}
\mathcal{L}_n^{(\alpha)}(\eta)=
\sum_{j=0}^{n}\frac{\Gamma(n+\alpha+1)}{\Gamma(j+\alpha+1)}\frac{(-\eta)^{j}}{j!(n-j)!}\,
\end{equation} 
are the generalized Laguerre polynomials.

On the other hand, application of the operator
$\hat{S}_{\nu}$ to the eigenstates (\ref{Calogerostates})
of the system $\hat{H}_\nu$ 
gives
\begin{eqnarray}
\label{coherent0}
&\hat{S}_{\nu}\psi_{\kappa,\nu}(\frac{1}{\sqrt{2}}x)=2^{\frac{1}{4}}e^{-\frac{1}{2}x^2+\frac{1}{4}\kappa^2}
\sqrt{x}\mathcal{J}_{\nu+1/2}(\kappa x):=\phi_\nu(x,\kappa)\,.&
\end{eqnarray} 
According to the first relation in (\ref{Dirac2+}),
these are the coherent states of the AFF model \cite{Perelomov},
\begin{equation}
\hat{J}_-\phi_\nu(x,\kappa)
=-\frac{1}{4}\kappa^2\phi_\nu(x,\kappa)\,.
\end{equation}
By allowing  the $\kappa>0$ to become a  complex parameter $z$,
coherent states can be constructed with  complex eigenvalues
of the operator $\hat{J}_-$.
Application of the evolution operator $\exp{-it\hat{H}^{\text{AFF}}_\nu}$ 
to these states gives the time-dependent coherent states
\begin{eqnarray}
\phi_{\nu}(x,z,t)=2^{1/4}\sqrt{x}\mathcal{J}_{\nu+1/2}(z(t) x)e^{-x^2/2+z^2(t)/4-it}\,,
\end{eqnarray} 
where $z(t)=z e^{-it}$.
In the case of $\nu=0$, these time-dependent coherent states of 
the AFF
model are the odd Schr\"odinger cat states of the
quantum harmonic oscillator \cite{Cat states}, 
\begin{equation}
\phi_{0}(x,z,t)\propto e^{-\frac{x^2}{2}+\frac{z^2(t)}{4}-\frac{it}{2}}\sin(z(t)x)\,.
\end{equation}

Similar to the analysis presented in the previous section, one can consider 
the classical picture underlying the quantum nonunitary  similarity transformation
associated with the constructed conformal bridge between the systems (iii) and (iv).
We will not discuss this picture in detail here, and only note that 
instead of (\ref{Tabg0}), we will have the phase space function of a similar form but
with $H_0$ changed for the Hamiltonian of the classical two-particle
Calogero system, $H_\nu=\frac{1}{2}p^2+\Delta_\nu$.
The phase space functions $H_\nu$, $D$ and $K$ satisfy the classical  algebra 
of the form (\ref{H0JD}), and the Casimir element $C_\nu=J^\mu J_\mu=-H_\nu^{\text{AFF}} K+D^2$
takes  here the value defined by the coupling constant,  $C_\nu=-\frac{1}{4}\nu(\nu+1)$.
Analogous to the pair discussed in Sec.\ref{SecClas0},
classical relations 
$(x^2)':=\{x^2,2iD\}=2ix^2$, $(p^2)':=\{p^2,2iD\}=-2ip^2$
for the  two-particle Calogero system
correspond here to the dynamics of the classical AFF model given by the relations
$\frac{d}{dt}J_\pm=\{J^\pm, H^{\text{AFF}}_\nu\}=\pm 2iJ_\pm$,
where $J_\pm$ are the classical analogs of the operators $\hat{J}_\pm$ from
(\ref{Dirac2+}).
\vskip0.1cm 

On the  Hilbert space
of the AFF system, the infinite-dimensional 
unitary  irreducible representation of the 
$\mathfrak{sl}(2,\R)$ algebra of the discrete type series $\mathcal{D}^+_\alpha$ 
with $\alpha=\frac{1}{2}\nu+3/4$ is realized, in which 
the states $\psi_{\nu,n}^{\text{AFF}}$ from (\ref{statesLnu1})  are eigenstates
of the compact generator $\hat{J}_0$ with eigenvalues $j_{0,n}=\alpha+n$, 
$n=0,1,\ldots$, and the Casimir operator takes the value
$\hat{C}_\nu=\hat{J}^\mu\hat{J}_\mu=-\alpha(\alpha-1)=
\frac{3}{16}-\frac{1}{4}\nu(\nu+1)$ \cite{MPsl2,AndrAFF,Kitaev}.

\section{Two-dimensional examples}
 \label{2Dcase}

When applying  the transformation $\hat{\mathfrak{S}}$ to generators of 
the conformal algebra,
one (formally) need not care about the concrete  realization of
the $\hat{J}_\mu$, since 
 only the algebraic relations presented by Eqs.
(\ref{SIGMA}) and (\ref{Automor}) are used. 
In particular, one may consider  to higher dimensional examples, 
where the range of physical systems of interest  is greater. 
In this section we  generalize
our construction to relate the two-dimensional free particle  system
with  a planar isotropic harmonic oscillator 
 and the Landau problem.
\vskip0.1cm

To begin with,  consider the nonunitary operator 
\begin{equation}\label{Sxy}
\hat{\mathfrak{S}}(x,y)=\hat{\mathfrak{S}}(x)
\hat{\mathfrak{S}}(y)
\end{equation}
with $\hat{\mathfrak{S}}(x)$ 
and $\hat{\mathfrak{S}}(y)$ of the form  (\ref{U0KH+}). 
Via a similarity transformation, it produces a  map 
\begin{eqnarray}
\hat{\mathfrak{S}}(x,y): (x,\hat{p}_x,y,\hat{p}_y)\rightarrow
 (\hat{a}^+_x,-i\hat{a}^-_{x},\hat{a}^+_y,-i\hat{a}^-_{y})\,.
\end{eqnarray}
Then the two-dimensional free particle
characterized by the dynamical conformal symmetry 
with generators
\begin{eqnarray}
\label{2dimensionConf}
\hat{H}=\frac{1}{2}(\hat{p}_x^2+\hat{p}_y^2)\,,\qquad 
\hat{D}=\frac{1}{2}(x\hat{p}_x+y\hat{p}_y +1)\,,\qquad
\hat{K}=\frac{1}{2}(x^2+y^2)\,
\end{eqnarray}
 is related with the planar isotropic harmonic oscillator 
 and generators of its Newton-Hooke symmetry as follows: 
 \begin{eqnarray}
 \label{sl2R2d}
 \hat{\mathfrak{S}}(x,y)\hat{H}\hat{\mathfrak{S}}^{-1}(x,y)=
 -\frac{1}{2}((\hat{a}_x^-)^2+(\hat{a}_y^-)^2)=-\hat{\mathcal{J}}_-\,,\label{MJ-}\\
 \hat{\mathfrak{S}}(x,y)2i\hat{D}\hat{\mathfrak{S}}^{-1}(x,y)=\hat{a}_x^+\hat{a}_x^-+\hat{a}_y^+
 \hat{a}_y^-+1=\hat{H}_{iso}=2\hat{\mathcal{J}}_0\,,\label{Hiso}\\
 \hat{\mathfrak{S}}(x,y)\hat{K}\hat{\mathfrak{S}}^{-1}(x,y)=\frac{1}{2}((\hat{a}_x^+)^2+(\hat{a}_y^+)^2)=
 \hat{\mathcal{J}}_+\,,\label{MJ+}
 \end{eqnarray}
 where the operators 
 $\hat{\mathcal{J}}_0$ and $\hat{\mathcal{J}}_\pm$ satisfy the $\mathfrak{sl}(2,\R)$ algebra
 (\ref{J+J-J0}).
  Analogously  to the  one-dimensional case, 
  the stationary and coherent states of  
  $\hat{H}_{\text{iso}}$ are produced by $\hat{\mathfrak{S}}(x,y)$
  from the corresponding states of the two-dimensional free particle
  system,
  \begin{eqnarray}
  &\hat{\mathfrak{S}}(x,y)\left(\frac{x}{\sqrt{2}}\right)^{n}
  \left(\frac{y}{\sqrt{2}}\right)^{m}=2^{n+m+\frac{1}{2}}e^{-\frac{(x^2+y^2)}{2}}H_{n}(x)H_{m}(y)\,,&\\
  &\hat{\mathfrak{S}}(x,y)e^{\frac{i}{\sqrt{2}}(k_x x+ k_y y)}=
  \sqrt{2} e^{-\frac{(x^2+y^2)}{2}}e^{\frac{k_x^2+k_y^2}{4}}e^{i(k_x x+ k_y y)} \,.&
  \end{eqnarray}

The angular momentum operator, 
\be
\hat{M}=x\hat{p}_y-y\hat{p}_x=
-i(\hat{a}_x^\dagger \hat{a}^-_y- \hat{a}_y^\dagger \hat{a}^-_x)\,,
\ee
is an integral of the 
planar free particle and of the isotropic harmonic oscillator systems.
It is invariant under the similarity transformation
produced by $\hat{\mathfrak{S}}$, or,  equivalently,
\begin{equation}
\hat{\mathfrak{S}}(x,y)\hat{M}=\hat{M}\hat{\mathfrak{S}}(x,y)\,. 
\end{equation}

Consider now the Landau problem for a scalar particle on $\R^2$.
In the symmetric gauge
$\vec{A}=\frac{1}{2}B(-q_2,q_1)$, 
the Hamiltonian operator (in units $c=m=\hbar=1$), 
\begin{equation}
\label{Landau}
\hat{H}_{\text{L}}=\frac{1}{2}\hat{\vec{\Pi}}^{2},\qquad
\hat{\Pi}_j=-i\frac{\partial}{\partial q_j}-eA_j\,, \qquad
[\hat{\Pi}_1,\hat{\Pi}_2]=ieB\,,
\end{equation}
has an explicitly rotational  invariant form.
Assuming $\omega_c=eB>0$, it can be  factorized,
\begin{eqnarray}
&\hat{H}_{\text{L}}=\omega_c(\hat{\mathcal{A}}^+\hat{\mathcal{A}}^-+\frac{1}{2})\,, &
\end{eqnarray}
 in terms of the 
Hermitian conjugate operators  
\begin{eqnarray}
\label{Landaulad}
\hat{\mathcal{A}}^\pm=\frac{1}{\sqrt{2\omega_c}}(\hat{\Pi}_1\mp
 i\hat{\Pi}_2)\,,\qquad [\hat{\mathcal{A}}^-,
\hat{\mathcal{A}}^+]= 1\,.
\end{eqnarray}
Setting  $\omega_c=2$, we can identify $q_i$ with dimensionless variables  $q_1=x$,
$q_2=y$. Then  we present $\hat{\mathcal{A}}^\pm$ as linear combinations of the 
usually defined ladder operators $\hat{a}^\pm_x$ and $\hat{a}^\pm_y$,
 in terms of which we also 
define the  operators $\hat{\mathcal{B}}^\pm$,
\begin{equation}
\hat{\mathcal{A}}^\pm=\frac{1}{\sqrt{2}}
(\hat{a}_y^\pm \pm i\hat{a}_x^\pm)\,,\qquad
\hat{\mathcal{B}}^\pm=\frac{1}{\sqrt{2}}(\hat{a}_y^\pm \mp i\hat{a}_x^\pm)\,.
\end{equation}
The operators  $\hat{\mathcal{B}}^\pm$  satisfy relation 
$[\hat{\mathcal{B}}^-,\hat{\mathcal{B}}^+]=1$
and commute with $\hat{\mathcal{A}}^\pm$. They are integrals of motion, 
and their  noncommuting Hermitian linear  combinations 
$\hat{\mathcal{B}}^++\hat{\mathcal{B}}^-$
and  $i(\hat{\mathcal{B}}^+-\hat{\mathcal{B}}^-)$
correspond to coordinates of the center of the cyclotron motion.
 In terms of the ladder operators 
$\hat{a}^\pm_x$, $\hat{a}^\pm_y$ the Hamiltonian $\hat{H}_{\text{L}}$ takes the
form of a linear combination of the Hamiltonian of the isotropic oscillator and 
angular momentum operator,
\begin{equation}\label{HLM3iso}
\hat{H}_{\text{L}}=\hat{H}_{\text{iso}} -\hat{M}\,. 
\end{equation}
On the other hand, 
the angular momentum operator and isotropic oscillator's Hamiltonian 
are presented in terms of $\hat{\mathcal{A}}^\pm$ and 
$\hat{\mathcal{B}}^\pm$
as follows,
\begin{equation}\label{M3Hiso}
\hat{M}=
\hat{\mathcal{B}}^+\hat{\mathcal{B}}^--
\hat{\mathcal{A}}^+\hat{\mathcal{A}}^-\,,\qquad
\hat{H}_{\text{iso}}=\hat{\mathcal{B}}^+\hat{\mathcal{B}}^-+
\hat{\mathcal{A}}^+\hat{\mathcal{A}}^-+1\,,
\end{equation}
and we have the commutation relations
$
[\hat{M},\hat{\mathcal{B}}^\pm]=\pm\hat{\mathcal{B}}^\pm,$
$
[\hat{M},\hat{\mathcal{A}}^\pm]=\mp\hat{\mathcal{A}}^\pm.$ 
By taking into account the invariance of the angular momentum under
similarity transformation, we find that its linear combination with the dilatation operator 
is transformed into the Hamiltonian of the Landau problem,
\be
\hat{\mathfrak{S}}(x,y)(2i\hat{D}-\hat{M})\hat{\mathfrak{S}}^{-1}(x,y)=\hat{H}_{\text{L}}\,.
\ee

Let us now introduce a complex coordinate in the plane,
\begin{equation}
w=\frac{1}{\sqrt{2}}(y+ix)\,, \quad \text{and}\qquad \bar{w}\,.
\end{equation}
The elements of conformal algebra and angular momentum 
operator take then the form 
\begin{equation}
\label{conformalcomplex}
\hat{H}=-\frac{\partial^2}{\partial w \partial \bar{w}}\,,\quad
\hat{D}=\frac{1}{2i}\left(w\frac{\partial}{\partial w}+\bar{w}\frac{\partial}{\partial \bar{w}}+1\right)\,,\quad
\hat{K}=w\bar{w}\,,\quad
\hat{M}=\bar{w}\frac{\partial}{\partial \bar{w}}-w\frac{\partial}{\partial w}\,, 
\end{equation}
and  we find  that the operator (\ref{Sxy}) generates the  
similarity transformations 
\begin{eqnarray}
&
\hat{\mathfrak{S}}(x,y)w\hat{\mathfrak{S}}^{-1}(x,y)=\hat{\mathcal{A}}^+\,, 
\qquad
\hat{\mathfrak{S}}(x,y)\left(\frac{\partial}{\partial w}\right)
\hat{\mathfrak{S}}^{-1}(x,y)=\hat{\mathcal{A}}^-\,,
&\label{wASig}\\
&
\hat{\mathfrak{S}}(x,y)\bar{w}\hat{\mathfrak{S}}^{-1}(x,y)=\hat{\mathcal{B}}^+ \,,
\qquad
\hat{\mathfrak{S}}(x,y)\left(\frac{\partial}{
\partial \bar{w}}\right)\hat{\mathfrak{S}}^{-1}(x,y)=\hat{\mathcal{B}}^-\,,
&\label{wBSig}\\
&
\hat{\mathfrak{S}}(x,y)\left(w\frac{\partial} {\partial w}\right)\hat{\mathfrak{S}}^{-1}(x,y)=
\hat{\mathcal{A}}^+\hat{\mathcal{A}}^-\,,\qquad
\hat{\mathfrak{S}}(x,y)\left(\bar{w}\frac {\partial} {\partial \bar{w}}\right)\hat{\mathfrak{S}}^{-1}(x,y)
=\hat{\mathcal{B}}^+\hat{\mathcal{B}}^-\,.
&\label{wdwSig}
\end{eqnarray}
Observe that each pair of relations in (\ref{wASig}) and  (\ref{wBSig}) 
has a form similar to 
the  one-dimensional transformation (\ref{sigmaxp}),
where, however, the coordinate and momentum are Hermitian operators.

Simultaneous eigenstates of the operators $w\frac{\partial }{\partial w}$ and 
$\bar{w}\frac{\partial }{\partial\bar{w}}$,
which satisfy the relations 
$w\frac{\partial}{\partial w}\phi_{n,m}=n\phi_{n,m}$ 
and 
$\bar{w}\frac{\partial }{\partial\bar{w}}\phi_{n,m}=m\phi_{n,m}$
with $n,m=0,1,\ldots$, 
are
\begin{equation}
\phi_{n,m}(x,y)= w^n(\bar{w})^{m}=2^{-(n+m)/2}\sum_{k=0}^{n}
\sum_{l=0}^{m}{n\choose k}{m\choose l}(i)^{n-m+l-k}y^{k+l}x^{n+m-k-l}\,, 
\end{equation}
where the binomial theorem has been used.
Employing Eq. (\ref{conformalcomplex}) we find that 
\begin{eqnarray}
&
\hat{M}\phi_{n,m}=(m-n)\phi_{n,m}\,,\qquad
2i\hat{D}\phi_{n,m}=(n+m+1)\phi_{n,m}\,,&\\&
\label{KyHenphi}
\hat{K}\phi_{n,m}=\phi_{n+1,m+1}\,,\qquad
\hat{H}\phi_{n,m}=-nm\phi_{n-1,m-1}\,.
&
\end{eqnarray}
The last equality shows  
that  $\phi_{0,m}$ and $\phi_{n,0}$ are the 
zero energy eigenstates of the two-dimensional free particle, 
while the $\phi_{n,m}$ with $n,m>0$ are the  Jordan states
corresponding to the same zero energy value.
Application of the  operator $\hat{\mathfrak{S}}(x,y)$ 
to these functions yields
\begin{equation}
\label{landaustates}
\hat{\mathfrak{S}}(x,y)\phi_{n,m}(x,y)=2^{2(n+m)+\frac{1}{2}}e^{-\frac{(x^2+y^2)}{2}}H_{n,m}(y,x)
=\psi_{n,m}(x,y)\,,
\end{equation}
where 
\begin{equation}
H_{n,m}(y,x)=2^{-(n+m)}\sum_{k=0}^{n}
\sum_{l=0}^{m}{n\choose k}{m\choose l}(i)^{n-m+l-k}H_{k+l}(y)H_{n+m-k-l}(x)\,,
\end{equation}
are the complex Hermite polynomials, see \cite{Hermite}. 
These functions are  
eigenstates of the operators $\hat{H}_{\text{L}}$, $\hat{M}$
and $\hat{H}_{\text{iso}}$, 
\begin{eqnarray}
&
\hat{H}_{\text{L}}\psi_{n,m}=(n+\frac{1}{2})\psi_{n,m}\,,\qquad 
\hat{M}\psi_{n,m}=(m-n)\psi_{n,m}\,,&\\&
\hat{H}_{\text{iso}}\psi_{n,m}=(n+m+1)\psi_{n,m}\,,&
\end{eqnarray}
and we note that $\psi_{n,n}$ is rotational invariant.

Equations (\ref{wASig}), (\ref{wBSig}),  and (\ref{KyHenphi}) show that 
the operators $\hat{\mathcal{A}}^\pm$ and $\hat{\mathcal{B}}^{\pm}$ act as the ladder operators 
for the  indexes  $n$ and $m$, respectively, 
while the operators $\hat{\mathcal{J}}_\pm$ given by Eqs. (\ref{MJ-})
and (\ref{MJ+})
increase or decrease simultaneously
  $n$ and $m$ by one.

Application 
of the operator  $\hat{\mathfrak{S}}(x,y)$ to exponential functions of the most general 
form   $e^{\alpha w+\beta \bar{w}}$ with $\alpha,\beta \in \mathbb{C}$
 gives  here, similar to the one-dimensional case,  the coherent states
of the Landau problem as well of the isotropic harmonic oscillator,
\begin{eqnarray}
\label{coherentLandau}
\begin{array}{lcl}
\psi_{\text{L}}(x,y,\alpha,\beta)&=& \hat{\mathfrak{S}}(x,y)
e^{\frac{1}{\sqrt{2}}((\alpha+\beta)y +i(\alpha-\beta)x)}=\sqrt{2}
e^{-\frac{(x^2+y^2)}{2}+(\alpha+\beta)y +i(\alpha-\beta)x-\alpha\beta}\\
&=& \sum_{n=0}^{\infty}\sum_{l=0}^{n}\frac{1}{n!}{n \choose l}\alpha^{l}\beta^{n-l}\psi_{l,n-l}(x,y)\,.
\end{array}
\end{eqnarray}
Applying to them, in particular,  the evolution operator $e^{-it\hat{H}_{\text{L}}}$,
we  obtain the
time dependent solution to the Landau problem, 
\begin{equation}
\psi_{\text{L}}(x,y,\alpha,\beta,t)=e^{-\frac{i t}{2}}\psi_{\text{L}}(x,y,\alpha e^{-it},\beta)\,, 
\end{equation} 
whereas under rotations these states transform as 
\begin{equation}
e^{i\varphi \hat{M}}\psi_{\text{L}}(x,y,\alpha,\beta)=\psi_{\text{L}}(x,y,\alpha e^{-i \varphi},\beta e^{i \varphi})\,.
\end{equation}
As the function  $e^{\alpha w+\beta \bar{w}}$
is a common eigenstate of the differential operators
$\frac{\partial}{\partial w}$ and  $\frac{\partial}{\partial \bar{w}}$
with eigenvalues $\alpha$ and $\beta$, respectively, then our 
transformation yields 
\begin{equation}
\hat{\mathcal{A}}^-\psi_{\text{L}}(x,y,\alpha,\beta)=\alpha\psi_{\text{L}}(x,y,\alpha,\beta)\,,\qquad
\hat{\mathcal{B}}^-\psi_{\text{L}}(x,y,\alpha,\beta)=\beta\psi_{\text{L}}(x,y,\alpha,\beta)\,, 
\end{equation}
which provides another explanation  why the wave functions 
(\ref{coherentLandau})  are 
the coherent states  for the planar harmonic oscillator as well as
for the Landau problem. 

\section{Discussion and outlook} 
\label{Discussion}

The intertwining operators of the Darboux transformations
allow one to construct eigenstates of a Hamiltonian of one system 
in terms of eigenstates of a Hamiltonian of another, related system.
There, the intertwining operators are differential operators of finite order, and so
the relation between corresponding eigenstates is local.
Those operators also factorize the corresponding Hamiltonians 
or polynomials thereof.
In our ``conformal bridge" constructions, the 
generator of the similarity transformation
is a nonlocal operator, formally given by 
an infinite series in the momentum operator.
It has a nature of the fourth order root of the identity operator
from the point of view of its action on generators of the conformal algebra,
where it acts as a nonunitary automorphism.
Its peculiarity in comparison with generators
of the Darboux transformations is that it intertwines 
the generators of the $\mathfrak{sl}(2,\R)$ having 
a different topological nature
corresponding to a change of the  dynamics in the sense of
Dirac. In fact, our transformation changes 
the conformally invariant asymptotically free dynamics
described by the Hamiltonian being a 
noncompact generator of the $\mathfrak{sl}(2,\R)$
to the conformal Newton-Hooke dynamics generated by a 
compact  generator.

\vskip0.1cm
In comparison with the well-known Niederer's canonical transformation
 \cite{Nied,Steu},
\begin{equation}\label{Nied}
\zeta=\frac{x}{\sqrt{1+(\omega t)^2}}\,,\qquad
d\tau=\frac{dt}{1+(\omega t)^2}\,,
\end{equation}
employed to transform the one-dimensional 
conformal mechanics model
 into the regularized AFF model, 
 our quantum nonunitary similarity transformations 
 and corresponding complex classical canonical transformations are time independent:
 we work within  the stationary framework. 
 As a result, instead of the map between solutions of the time-dependent 
 Schr\"odinger equation of the free particle and harmonic oscillator,
 or of the conformal mechanics corresponding to the  two-particle 
 Calogero model and the AFF model, our transformation 
 acts in an  essentially different way.
 It maps formal eigenstates of the dilatation operator with imaginary 
 eigenvalues being also 
 Jordan states  corresponding to the zero energy value of the conformal mechanics model 
Hamiltonian $\hat{H}_\nu$ 
(or of the free particle Hamiltonian $\hat{H}_0$) into physical bound eigenstates of 
the regularized AFF model $\hat{H}_\nu^{\text{AFF}}$ (or of the  harmonic oscillator 
$\hat{H}_{\text{osc}}$)
of the real energy values.
Besides, the  stationary eigenstates  of $\hat{H}_\nu$ (or, $\hat{H}_0$) are transformed 
 into  the coherent states of $\hat{H}_\nu^{\text{AFF}}$ ($\hat{H}_{\text{osc}}$).

\vskip0.1cm
The interesting question is then what is a possible geometrical interpretation of our
complex canonical transformation,
being an
eighth order root of the identity transformation.
One could try to characterize it 
within the framework of the Eisenhart lift \cite{Eise}
or by using the Dirac trick  
which corresponds to
a generation of the time-dependent Schr\"odinger equation via the
presentation of the classical action in the reparametrization invariant form.
This way one may expect to establish a possible relation (if it exists) of our
``conformal bridge" constructions with the transformations based 
on the Niederer's time-dependent canonical transformation (\ref{Nied}).
Investigation of the indicated problem may be of interest in 
the light of  the AdS/CFT correspondence \cite{MiStr,BlackHold1,Jack}.

\vskip0.08cm
Our canonical transformations correspond to the Hamiltonian vector flows 
 produced by generators of the conformal symmetry 
 with  particular complex values of the parameters.
The interesting question is whether such transformations
having    
the nature of 
the 
fourth order root of the spatial reflection have any interpretation in the context
of  $\mathcal{PT}$ symmetry \cite{Bender,PTSUSY,PT}. This question is intriguing especially having in mind 
that perfectly invisible zero-gap systems and the related  extreme wave solutions (multi-solitons)  to the  
complexified KdV equation can be obtained from the free particle system
by application to it of the $\mathcal{PT}$-regularized Darboux transformations intimately related to conformal 
symmetry \cite{confDarb,JM,JM2}.

It is not difficult to find the analog of our canonical transformation   
that maps 
 the dilatation generator of the free particle
into the  Hamiltonian 
of the inverted (repulsive) harmonic oscillator
having  
a  continuous spectrum at the 
quantum level \cite{LL}.  
In this way, Gamow states 
\cite{Gamow}
enter naturally into the construction.
However, 
at the quantum level there appears a  problem because the transformation, 
 which now will be unitary, acting on the eigenstates of the dilatation  operator with real eigenvalues 
will  produce divergent series.  This   implies 
the necessity of a deeper analysis  of the corresponding unitary transformation.   
 
 \vskip0.08cm
 Yet another interesting question is whether  by using a conformal bridge 
 transformation,
 new solutions to some integrable systems in partial derivatives can be constructed. 
  We have here in mind
 the generation of the multisoliton solutions to the KdV equation,
 including rational solutions of the Calogero  type,
 by application of the Darboux transformations to the quantum free particle system
 on the one hand, and the conformal bridge relation of  the free particle with
 the quantum harmonic oscillator  that we have discussed. 
 Such a hypothetical possibility still remains a mystery for us.
 
 \vskip0.08cm   
As our transformation does not depend on a concrete  realization of the generators of the conformal algebra,
as we showed, one can  study higher dimensional systems.  In this way we found 
an interesting relation between the 2D free particle system and the Landau problem,
where the two-dimensional plane waves were  mapped into the coherent states of the
planar harmonic oscillator and of the Landau problem.
It would be interesting to investigate 
 what happens in the case of the  higher-dimensional nonseparable conformal invariant systems
such as  multiparticle Calogero model\footnote{Based on  a spectrum similarity, 
 Calogero conjectured  \cite{Calogero1}   that there should exist a  
nontrivial relation between his model 
and the system of decoupled oscillators. 
This conjecture was proved, including the case without quadratic  interactions, 
by constructing some (non)-unitary transformations that mutually map Hamiltonians of the 
corresponding systems of the coupled and decoupled particles 
\cite{Calogerofree1,Calogerofree2,Calogerofree3,Calogerofree4} and 
give rise to a non-local $\mathfrak{sl}(2,\R)$ generator \cite{Calogerofree4}.
The conformal transformations we discussed here, in spite of being 
of a somewhat similar form  to those in 
\cite{Calogerofree2,Calogerofree3,Calogerofree4}, 
are   considered from an essentially different   
perspective of the change of the form of dynamics \emph{\'a la} Dirac \cite{Dirac,AFF,MiStr,BlackHold1,holQCD}.}, 
or in the case of the Dirac magnetic monopole system.

\vskip0.1cm

Conformal bridge transformations that we considered are based on the global dynamical
$\mathfrak{sl}(2,\R)$ symmetry. 
This global symmetry is promoted to the local conformal  [symplectic,
$\mathfrak{sp}(2,\R)\cong \mathfrak{sl}(2,\R) $]
gauge symmetry in 2T (two-time) physics;
see \cite{Bars1,Bars2,PlyuLeiva,Bars3,Bonezzi1,Bonezzi2} and further references therein. 
The 2T physics provides a ``holographic" unification  of different 1T systems, which appear 
there  under different choices of gauge fixing  and are treated as different  ``shadows" 
of the extended system with gauged conformal  symmetry. It would be very interesting to 
consider the conformal bridge transformations in the light of the 2T physics approach and 
somehow related to it contact geometry and  Reebs dynamics 
\cite{Herczeg1,Herczeg2}. The particular interesting 
question in this context is how  the quantum Jordan states corresponding to zero 
energy would reveal themselves in the 2T physics and contact quantization,
 and in the associated duality relations. 
 Another interesting point  to clarify is how the complex nature of the generalized 
 canonical transformations, which underlie  the quantum conformal 
 bridge transformations, will manifest itself in 2T physics. 
The study of the conformal bridge transformations in the context of 2T physics
also 
is promising having in mind that the latter approach finds applications to a very 
broad class of the physical systems.

\vskip0.1cm
As a final interesting problem that seems  to be rather natural for further investigation
we indicate  a  generalization of our constructions to the superconformal case
\cite{FubRab,SCM5},
bearing, particularly, in mind the presence of the hidden superconformal symmetry 
in the quantum harmonic oscillator \cite{InzPlyHid}.
It is worthwhile  to note  that the topic of supersymmetry  
was also addressed  within the framework of  2T physics
\cite{Bars2,Bonezzi2}
and contact geometry 
\cite{Bruce}.  \vskip0.05cm

Some  results on the listed problems are presented in our paper
 \cite{IPW}.    

 \vskip0.2cm
{\bf  Acknowledgements}
\vskip0.1cm

The work was partially supported by the CONICYT scholarship 21170053 (LI),
FONDECYT Project 1190842 (MSP), by the Project USA 1899 (MSP and LI),
and  by DICYT,  USACH (MSP).
LI and MSP are grateful to Jena University for the warm hospitality
where a part of this work was realized. 
We thank O.~Lechtenfeld for drawing our attention to 
\cite{Calogerofree4} and related 
refs. \cite{Calogerofree1,Calogerofree2,Calogerofree3}.

\appendix

\section*{Appendix}

\section{Conformal symmetry}

The $\mathfrak{so}(2,1)\cong \mathfrak{sl}(2,\R)\cong \mathfrak{su}(1,1)
\mathfrak{sl}(2,\R)
   \cong \mathfrak{sp}(2,\R)
  $ 
algebra is given by the commutation relations
  \begin{equation}\label{Jmunul}
  [J_\mu,J_\nu]=-i\epsilon_{\mu\nu\lambda}J^\lambda\,,
  \end{equation}
 where $\epsilon^{\mu\nu\lambda}$ is a totally  antisymmetric tensor,
 $\epsilon^{012}=1$,
 and the metric  is $\eta^{\mu\nu}=\text{diag}\,(-1,1,1)$.
 The algebra (\ref{Jmunul}) can be  generated, in particular,  
  by  $J_\mu=-\epsilon_{\mu\nu\lambda}x^\nu p^\lambda$ 
 realized in terms of the $(2+1)$-dimensional 
 coordinate and momenta operators, $[x_\mu,p_\nu]=i\eta_{\mu\nu}$. 

 With 
 \begin{equation}\label{Jpmdef}
 J_\pm:=J_1\pm i J_2\,, 
 \end{equation}
 the algebra 
 (\ref{Jmunul}) 
  takes  the form of the conformal $\mathfrak{sl}(2,\R)$ algebra
   \begin{equation}\label{J+J-J0}
   [J_-,J_+]=2J_0\,,\qquad
   [J_0,J_\pm]=\pm J_\pm\,.
   \end{equation}
   The transformation
     \begin{equation}\label{automor}
     J_0\rightarrow
     -J_0\,,\qquad
     J_\pm\rightarrow -J_\mp
     \end{equation}
     defines an outer automorphism of (\ref{J+J-J0}), and
     the Casimir element of the algebra is 
     \begin{equation}\label{Casimir}
     C=J_\mu J^\mu=-(J_0)^2+(J_1)^2+(J_2)^2
     =-(J_0)^2+\frac{1}{2}(J_+J_-+J_-J_+)\,.
     \end{equation}

   The realization 
   \begin{equation}\label{Generators-t}
    J_-=\frac{\partial}{\partial t}\,,\qquad
   J_0=t\frac{\partial}{\partial t}\,, \qquad
J_+=t^2\frac{\partial}{\partial t}
\end{equation}
  generates the  conformal transformations 
   \begin{equation}\label{transf-t}
   t\rightarrow t+\alpha\,, \qquad
     t\rightarrow e^{\beta}t\,,\qquad
      t\rightarrow \frac{t}{1-\gamma t}
      \end{equation}
      of a time variable.
The    transformations (\ref{transf-t})
      together with 
      \begin{equation}
      x\rightarrow  e^{\frac{1}{2}\beta}x\,,\qquad
      x\rightarrow \frac{x}{1-\gamma t}
      \end{equation}
  are generated by
        \begin{equation}\label{Generators-xt}
    J_-=\frac{\partial}{\partial t}\,,\qquad
   J_0=t\frac{\partial}{\partial t}+\frac{1}{2} x\frac{\partial}{\partial x}\,, \qquad
J_+=t^2\frac{\partial}{\partial t}+xt\frac{\partial}{\partial x}\,,
\end{equation}
and represent the  dynamical conformal symmetry of the
      free particle   action  $\mathcal{A}=\frac{m}{2}\int \dot{x}^2 dt$.
  
      The formal change $t\rightarrow z$, $z\in \C$,  in (\ref{Generators-t}) 
      accompanied by the automorphism 
      (\ref{automor}) with subsequent identification $L_0=J_0$,
      $L_{+1}=J_-$, $L_{-1}=J_+$  gives generators of the $\mathfrak{su}(1,1)$ subalgebra 
      of the Virasoro algebra, 
        \begin{equation}
        L_0=-z\frac{\partial}{\partial z}\,,\qquad
        L_{+1}=-z^2\frac{\partial}{\partial z}\,,\qquad
        L_{-1}=-\frac{\partial}{\partial z}\,.
        \end{equation}
     This can be realized as
        \begin{equation}
        J_0=a^+a^-\,,\qquad
        J_+=(a^+)^2a^-\,,\qquad
        J_-=a^-\,
        \end{equation} 
   with ladder operators 
   $a^\mp=\frac{1}{\sqrt{2}}(x\pm\frac{d}{dx})$
   of the quantum harmonic oscillator.
   Another realization of (\ref{J+J-J0}) is given by
   \begin{equation}\label{J+J-J-0QHO}
   J_0=\frac{1}{4}(a^+a^-+a^-a^+)\,,\qquad
   J_{\pm}=\frac{1}{2}(a^\pm)^2\,,
   \end{equation}
   which corresponds to the conformal Newton-Hooke symmetry of the quantum harmonic oscillator.

\section{Hamiltonian vector flows as canonical transformations}\label{AppB}

A Hamiltonian vector flow 
generated  by a function $F$ on a  phase space $\mathcal{M}$
is given by 
\begin{equation}\label{TransCan}
f(\alpha)=\exp(\alpha F)\star f(q,p):=f(q,p)+\sum_{n=1}^\infty 
\frac{\alpha^n}{n!}\{F,\{\ldots,\{F,f\underbrace{\}\ldots\}\}}_{n}=:T_F(\alpha)(f)\,.
\end{equation}
The parameter  $\alpha$ is usually assumed to be real, but we allow for 
complex values.
Transformations (\ref{TransCan}) correspond 
to the action of a one-parametric Lie group on $\mathcal{M}$,
$$T_F(\alpha)\circ T_F(\beta)=T_F(\alpha+\beta),\qquad
 T_F(0)=I,\qquad
(T_F(\alpha))^{-1}=T_F(-\alpha).
$$
The composition of the Hamiltonian flows generated by functions $F$ and $G$ 
with $\{F,G\}\neq 0$ is noncommutative, and 
$$
(T_F(\alpha)\circ T_G(\beta))^{-1}=T_G(-\beta)\circ T_F(-\alpha).
$$
For functions $f$ and $g$ on phase space, 
the following  relation holds
\begin{equation}\label{Tfg}
T_F(\alpha)\circ T_G(\beta)(f\cdot g)=(T_F(\alpha)\circ T_G(\beta)(f))\cdot(T_F(\alpha)\circ T_G(\beta)(g)).
\end{equation}
A flow of a Hamiltonian vector field is a canonical transformation\,:
$\{f(\alpha),g(\alpha)\}=\{f,g\}$. 
In the general case of $\alpha\in \C$, the transformation 
(\ref{TransCan}) corresponds to the quantum similarity transformation
$\hat{f}(\alpha)=\exp(-i\alpha\hat{F})\hat{f}\exp(i\alpha\hat{F})$.

\end{document}